\documentstyle[aps,prb,multicol,epsf]{revtex}

\begin{document}

\title{
Spinless impurities and Kondo-like behavior
in strongly correlated electron systems
}

\author{Satoshi Fujimoto}
\address{
Department of Physics,
Kyoto University, Kyoto 606, Japan
}

\date{\today}
\maketitle
\begin{abstract}
We investigate magnetic properties induced by a spinless impurity
in strongly correlated electron systems, i.e.
the Hubbard model in the spatial dimension $D=1,2,$ and $3$.
For the 1D system exploiting the Bethe ansatz exact solution 
we find that the spin susceptibility and the local density of states 
in the vicinity of a spinless impurity
show divergent behaviors. The results imply that 
the induced local moment is not completely quenched 
at any finite temperatures.
On the other hand, the spin lattice relaxation rate 
obtained by bosonization and boundary conformal field theory
satisfies a relation analogous to the Korringa law, $1/T_1T \sim \chi^2$.
In the 2D and 3D systems, the analysis based upon the antiferromagnetically
correlated Fermi liquid theory reveals that
the antiferromagnetic spin fluctuation developed in the bulk
is much suppressed in the vicinity of a spinless impurity, and thus
magnetic properties are governed by the induced local moment, which leads to
the Korringa law of $1/T_1$.
\end{abstract}


\section{Introduction}

Recently, magnetic properties induced by spinless impurities
in correlated electron systems have attracted 
much interest.\cite{nmr1,nmr2,nmr3,nmr4,voj1,voj2}
Especially, to probe antiferromagnetic correlations of High-$T_c$ cuprates
the substitution of Cu sites with non-magnetic impurities such as
Zn, Al, and Li, has been studied experimentally.\cite{nmr1,nmr2,nmr3,nmr4}
According to NMR experiments, it was found that the substitution with
spinless impurities induces local moments in the vicinity of
impurities, which also show Kondo-like behaviors.
For instance, the spin susceptibility in the vicinity of an impurity
shows the temperature dependence like $\sim 1/(T+T_{\rm K})$, which
 implies the existence of the characteristic energy scale $T_{\rm K}$
analogous to Kondo temperature.\cite{nmr3}
Moreover the spin lattice relaxation rate $1/T_1$ shows Korringa-like
behaviors, $1/T_1T \propto K^2$, for $T< T_{\rm K}$.\cite{nmr4} 
Here $K$ is the Knight shift.
It is noted that in the vicinity of a spinless impurity
the antiferromagnetic spin correlation which is developed
in the bulk is much suppressed, and the magnetic correlation
is dominated by the induced local moment. From theoretical 
points of view, it is non-trivial how
this induced local moment governs the magnetic properties around
an impurity, suppressing the antiferromagnetic correlation. 
In this paper, we shall deal with this issue.
Although the experiments are carried out for High-$T_c$ cuprates which
are essentially quasi-two-dimensional systems, it is expected that
such effects may depend on the lattice structure and the dimensionality.
Thus, we consider the Hubbard models with a spinless impurity
in the spatial dimension $D=1,2,$ and $3$ to investigate how
the dimensionality affects the induced magnetic properties.
For $D=1$, the effects of a spinless impurity is incorporated
into an open boundary condition as will be explained in the next section.
Thus we consider the 1D Hubbard model with boundaries which
is exactly solvable in terms of the Bethe ansatz method.
We analyze the magnetic properties of this model using the exact solution 
and boundary conformal field theory.
For $D=2$ and $3$, we derive the Korringa relation satisfied
in the vicinity of a spinless impurity which is observed
in NMR experiments.
Our argument for $D=2$, $3$ is based upon the Fermi liquid theory 
in the presence of antiferromagnetic spin fluctuations.

The organization of this paper is as follows.
In Sec. II, the 1D Hubbard model with a spinless impurity
is considered.
The spin susceptibility and the local density of states
in the vicinity of an impurity
are obtained based upon the Bethe ansatz exact solution.
It is found that the induced moment is not screened completely
at any finite temperatures.
We also derive the spin-lattice relaxation rate $1/T_1$
which satisfies a relation analogous to the Korringa law.
In Sec. III, we discuss about the 2D and 3D systems exploiting
the antiferromagnetically correlated Fermi liquid theory.
Summary is given in Sec. IV.

\section{A spinless impurity in the one-dimensional Hubbard model}

\subsection{Mapping to the Hubbard model with boundaries 
and the Bethe ansatz exact solution}

The effects of a single impurity 
in one-dimensional correlated systems have been extensively studied 
so far.\cite{kane,furu}
If the interaction between fermions is repulsive,
a potential scattering in the 1D Hubbard model
is renormalized to an infinite strength, eventually,
cutting the system into two half-infinite chains 
in the low-energy scaling limit.  
Thus at sufficiently low temperatures the system can be treated as
the Hubbard chain with open boundaries, of which
the hamiltonian is given by,
\begin{equation}
H=-\sum_{\sigma, i=1}^{L-1}c^{\dagger}_{\sigma i}c_{\sigma i+1}+h.c.
+U\sum_{i=1}^{L}n_{\uparrow i}n_{\downarrow i}
-\mu\sum_{\sigma, i=1}^{L}n_{\sigma i}
-\frac{H}{2}\sum_{i=1}^{L}(n_{\uparrow i}-n_{\downarrow i})
+V\sum_{\sigma}n_{\sigma 1}, \label{hamil}
\end{equation}
where the last term is a boundary potential.
As we will see below, the low-energy spin dynamics around the impurity
with which we are concerned are mainly described by this model,
and the interaction or hopping between the two half-infinite chains
is a subleading irrelevant interaction which can be incorporated
by perturbative calculation.

The Bethe ansatz exact solutions of 1D correlated systems with boundaries 
have been studied by many 
authors.\cite{gau,hschu,woy,alca,yam1,desa,fuji,ess,asa,fra,yam,degu,shiro}
In connection with the spin dynamics in the vicinity of the boundary,
an intriguing result was obtained for the supersymmetric {\it t-J} model
by Essler.\cite{ess} 
He obtained the divergent behavior of the boundary spin susceptibility
as a function of a magnetic field $H$, i.e.
 $\chi_{\rm boundary} \sim 1/H(\ln H)^2$.
It was first predicted by de Sa and Tsvelik that such a Curie-like
behavior is universal for integral models with boundaries.\cite{desa}
Later, the similar behavior was also found 
for the Hubbard model at half-filling by Asakawa and Suzuki.\cite{asa}
In the next subsection, we shall show that this divergent behavior
holds also for the case away from half-filling with finite $U$.

Here we summarize the basic equations which are
relevant to the following arguments.
The Bethe ansatz equations of the 1D Hubbard model with boundaries
obtained by Schulz many years ago are,\cite{hschu}
\begin{eqnarray}
e^{i2k_jL}e^{i\phi_0(k_j)}&=&
\prod_{\beta=1}^{M}e_1(\sin k_j-\lambda_{\beta}) 
e_1(\sin k_j+\lambda_{\beta}), \label{bethe1} \\
\prod_{j=1}^N e_1(\lambda_{\alpha}-\sin k_j)
e_1(\lambda_{\alpha}+\sin k_j)&=& 
\prod_{{\scriptstyle \beta=1}\atop{\scriptstyle \beta\neq\alpha}}
e_2(\lambda_{\alpha}-\lambda_{\beta}) 
 e_2(\lambda_{\alpha}+\lambda_{\beta}), \label{bethe2}
\end{eqnarray}
where $e_{n}(x)=\frac{x+{\rm i}n u}{x-{\rm i}n u}$, $u=U/4$, and 
$\phi_{0,L}$ is a potential  at boundaries.
$N$ is the total number of electrons. $M$ is the total number of down spins.
$k_j$ and $\lambda_{\alpha}$ are rapidities for charge and spin degrees
of freedom, respectively.
In the following, we consider only the case of repulsive 
boundary potentials.
Thus the above equations have real roots.
Putting $k_{-j}=-k_j$, $\lambda_{-\alpha}=-\lambda_{\alpha}$,
and taking a continuum limit, we have the integral equations
for the distribution functions of rapidities,
\begin{eqnarray}
\rho(k)&=&\frac{1}{\pi}+\frac{1}{\pi L}\phi_0'(k)
-\frac{1}{2\pi L}\frac{2u\cos k}{(\sin k)^2+u^2}
+\cos k \int^{B}_{-B}\frac{d\lambda}{\pi}
\frac{u}{(\sin k-\lambda)^2+u^2}\sigma(\lambda), \label{rho} \\
\sigma(\lambda)&=&\frac{1}{\pi L}\frac{2u}{\lambda^2+4u^2}
+\int^{Q}_{-Q}\frac{dk}{\pi}\frac{u}{(\lambda-\sin k)^2+u^2}\rho(k)
-\int^{B}_{-B}\frac{d\lambda'}{\pi}\frac{2u}{(\lambda-\lambda')^2+4u^2}
\sigma(\lambda'). \label{sigma}
\end{eqnarray}
$N$ and $M$ are given by,
\begin{equation}
\int^{Q}_{-Q}\rho(k)dk=\frac{2N+1}{L}, \label{cha}
\end{equation}
\begin{equation}
\int^B_{-B}\sigma(\lambda)d\lambda=\frac{2M+1}{L}.  \label{spin}
\end{equation}
Then the magnetization is expressed as,
\begin{equation}
\frac{S_z}{L}=\frac{1}{4}\int^{Q}_{-Q}\rho(k)dk-\frac{1}{2}\int^B_{-B}
\sigma(\lambda)d\lambda+\frac{1}{4L}. \label{spinz}
\end{equation}
The total energy is expressed in terms of the dressed energies,
\begin{equation}
\frac{E}{L}=\int^{Q}_{-Q}dk\biggl(\frac{1}{\pi}+\frac{1}{\pi L}\phi_0'(k)
-\frac{1}{2\pi L}\frac{2u\cos k}{(\sin k)^2+u^2}\biggr)\varepsilon_c(k)
+\int^{B}_{-B}\frac{d\lambda}{\pi L}\frac{2u}{\lambda^2+4u^2}
\varepsilon_s(\lambda),  \label{energy}
\end{equation}
where the dressed energies $\varepsilon_c(k)$ and $\varepsilon_s(\lambda)$
are determined by the integral equations,
\begin{eqnarray}
\varepsilon_c(k)&=&-2\cos k -\frac{H}{2}-\mu 
+\int^B_{-B}\frac{d\lambda}{\pi}\frac{u}{(\sin k-\lambda)^2+u^2}
\varepsilon_s(\lambda), \label{enech} \\
\varepsilon_s(\lambda)&=&H+\int^{Q}_{-Q}\frac{dk}{\pi}
\frac{u}{(\sin k-\lambda)^2+u^2}\varepsilon_c(k)
-\int^B_{-B}\frac{d\lambda'}{\pi}\frac{2u}{(\lambda-\lambda')^2+4u^2}
\varepsilon_s(\lambda'). \label{enespin}
\end{eqnarray}
If one fixes the magnetic field $H$,
$B$ is determined by the equilibrium condition $\partial E/\partial B=0$,
which is equivalent to the condition, $\varepsilon_s(B)=0$.
In the subsequent sections, we calculate the spin susceptibility
and the local density of states using the above equations.

\subsection{Spin susceptibility}

In order to derive the spin susceptibility, we solve 
eqs.(\ref{rho}) and (\ref{sigma}) for $\sigma(\lambda)$
using the Wiener-Hopf method, 
and obtain the magnetization, eq.(\ref{spinz}). 

Applying the Fourier transformation and shifting the argument,
$\lambda\rightarrow\lambda+B$, we rewrite eq.(\ref{sigma})
into,
\begin{eqnarray}
\sigma(\lambda+B)=f_0(\lambda+B)
+\int^{\infty}_0\frac{d\lambda'}{\pi}R(\lambda-\lambda')\sigma(\lambda'+B)
+\int^{\infty}_0\frac{d\lambda'}{\pi}
R(\lambda+\lambda'+2B)\sigma(\lambda'+B), \label{sigma2}
\end{eqnarray}
where 
\begin{eqnarray}
f_0(\lambda+B)&=&\frac{1}{L}\int^{\infty}_{-\infty}\frac{d\omega}{2\pi}
\frac{e^{-u\vert \omega\vert}e^{-i\omega(\lambda+B)}}{2\cosh u\omega}
+\int^{Q}_{-Q}dk\frac{\rho(k)}
{2\cosh \frac{\pi}{u}(\lambda+B-\sin k)}, \\
R(x)&=&\int^{\infty}_{-\infty}\frac{d\omega}{2\pi}
\frac{e^{-u\vert \omega\vert}e^{-i\omega x}}{2\cosh u\omega}.
\end{eqnarray}
The last term of the right-hand side of eq.(\ref{sigma2})
is $O(1/B^2)$ for small magnetic fields.
Thus we neglect it.
Then eq.(\ref{sigma2}) can be solved by using 
the standard Wiener-Hopf method.\cite{wh}
The solution is expressed in terms of the following functions,
\begin{eqnarray}
&&G^{+}(\omega)=\sqrt{2\pi}
\frac{(-i\frac{u\omega}{\pi})^{-i\frac{u\omega}{\pi}}}
{\Gamma(\frac{1}{2}-i\frac{u\omega}{\pi})}
e^{\frac{iu\omega}{\pi}}, \\
&&G^{-}(\omega)=(G^{+}(-\omega))^{-1}, \\
&&Q^{+}(\omega)+Q^{-}(\omega)=G^{-}(\omega)\tilde{f}_0(\omega), \label{qq}\\
&&\tilde{f}_0(\omega)= \int^{\infty}_{-\infty}
d\lambda f_0(\lambda+B)e^{i\omega\lambda},
\end{eqnarray}  
where $Q^{+}(\omega)$ ($Q^{-}(\omega)$) is the analytic part 
of $G^{-}(\omega)\tilde{f}_0(\omega)$ defined in the upper (lower)
half plane.
Fourier transforming eq.(\ref{sigma2}) and introducing
the function $\sigma^{+}(\omega)=\int^{\infty}_0d\lambda
e^{i\omega\lambda}\sigma(\lambda+B)$, we obtain the solution as,
$\sigma^{+}(\omega)=G^{+}(\omega)Q^{+}(\omega)$.

Now we derive $Q^{+}(\omega)$ as follows.
For small magnetic fields, i.e. large $B$, and $\lambda >0$
the second term of $f_0(\lambda+B)$ is approximated as,
\begin{equation}
\int^{Q}_{-Q}dk\frac{\rho(k)}
{2\cosh \frac{\pi}{u}(\lambda+B-\sin k)}\approx 
\frac{2N+1}{L}\frac{1}{2\cosh \frac{\pi}{u}(\lambda+B)}.
\end{equation}
This driving term is essentially the same as the bulk contribution,
with which we are not concerned.
The first term of $f_0(\lambda+B)$ gives rise an interesting 
boundary effect. 
Using $e^{-u\vert \omega\vert}/2\cosh u\omega =
\sum_{n=1}^{\infty}(-1)^{n-1}e^{-2n u \vert\omega\vert}$ and
the Laplace transformation,
\begin{equation}
\frac{2nu}{(\lambda+B)^2+(2nu)^2}=\int^{\infty}_0dt
e^{-(\lambda+B)t}\sin (2nut),
\end{equation}
we rewrite the first term of $f_0(\lambda+B)$ as,
\begin{equation}
\frac{1}{L}\int^{\infty}_{-\infty}\frac{d\omega}{2\pi}
\frac{e^{-u\vert \omega\vert}e^{-i\omega(\lambda+B)}}{2\cosh u\omega}
=\frac{1}{\pi L}\sum_{n=1}^{\infty}(-1)^{n-1}
\int^{\infty}_{-\infty}\frac{d\omega}{2\pi}
\int^{\infty}_0dt \sin(2nut)\biggl[\frac{1}{\omega+it}-\frac{1}{\omega-it}
\biggr]
ie^{-i\omega(\lambda+B)}. \label{ff0}
\end{equation}
The analytic property of eq.(\ref{ff0}) solves eq.(\ref{qq}),
\begin{equation}
Q^{+}(\omega)=\frac{1}{L}\sum_{n=1}^{\infty}\int^{\infty}_0dt
\sin (2nut)\frac{e^{-tB}}{\omega+it}iG^{-}(-it)+\mbox{bulk terms}.
\end{equation}
Finally, using eq.(\ref{spinz}), we obtain
the magnetization,
\begin{equation}
S_z=\frac{1}{2}\int^{\infty}_0d\lambda \sigma(\lambda+B)
=\frac{1}{2}\sigma^{+}(0)
\sim \frac{1}{LB} +\mbox{bulk terms},
\end{equation}
for large $B$, i.e. small magnetic fields.
$B$ is related to $H$ from the condition 
$\varepsilon_s(B)=0$. From eqs.(\ref{enech}) and (\ref{enespin}),
we have, $H=Ce^{-\pi B/2u}$ for $H\ll u$. Here $C$ is an constant.
Then the spin susceptibility $\chi=\partial S_z/\partial H$
behaves like, 
\begin{equation}
\chi\sim \frac{1}{L}\frac{1}{H(\ln H)^2}+\mbox{bulk terms}. \label{chi}
\end{equation}
This $H$-dependence is the same as that found 
for the half-filling case.\cite{asa}
The above result implies that in 1D systems 
the magnetic moment induced by a non-magnetic
impurity is not screened completely even at zero temperature.
This behavior is analogous to the underscreening multi-channel Kondo effect,
as pointed out by de Sa and Tsvelik.\cite{desa} 
The leading $H$-dependence of eq.(\ref{chi}) is not altered,
even if one includes irrelevant interactions such as
the hopping between the two half-infinite chains.

In this section, we restrict our discussion to the zero temperature case.
It is expected that at finite temperatures the boundary spin susceptibility
behaves like $\chi_{\rm boundary}\sim 1/T(\ln T)^2$.
In order to confirm this prediction, we need to explore
thermodynamic Bethe ansatz method in the presence of boundaries.
However, in the presence of boundaries, the entropy can not be
expressed in terms of rapidity distribution functions 
in the continuum limit, because of the presence of spurious states 
for vanishing rapidities, and thus the usual technique of thermodynamic Bethe 
ansatz method is not applicable.
If we limit the argument to sufficiently low temperature regions,
undesirable contributions from the spurious state around the bottom
of the energy spectrum may be small, and not give rise serious errors.
Even if we admit this approximation, it is still a cumbersome task to solve 
thermodynamic Bethe ansatz equations numerically for low temperatures.
Thus here we just give a field-theoretical argument
to justify the above speculation.
According to the boundary conformal field theory,
the above divergent behavior of the spin susceptibility
is due to the presence of a boundary entropy, 
$S_{\rm bound}=T\ln (\sqrt{4\pi}R)$.\cite{egg,asa}
Here $R$ is the radius of the boson field of the Gaussian model
which is the low-energy effective theory.
If the leading irrelevant interaction is the marginal operator 
in the spin degrees of freedom, $\bbox{J}_L\cdot \bbox{J}_R$,
we have $R\sim R_0-g/\ln T$ for small $T$.\cite{affgep} 
Then, the boundary spin susceptibility should behave like,
$\chi_{\rm boundary}\sim 1/T(\ln T)^2$.
Thus we expect that this temperature dependence which signifies
the presence of an unquenched local moment may realize
in this system.

\subsection{Local density of states}

In models solvable in terms of the Bethe ansatz method, 
the local density of states is defined as the derivative of
the quantum number, which parameterizes rapidities, 
with respect to the pseudo-energy, i.e. 
$\partial I_j/\partial \varepsilon(k_j)$.\cite{kawa}
For the 1D Hubbard model, we can consider the density of states 
of holon and spinon, respectively.
An interesting singular behavior due to the boundary 
appears in the spin degrees of freedom.

The local density of states of spinon as a function of energy is given by,
\begin{equation}
\rho_{\rm spin}(\varepsilon)=\frac{\partial \lambda}{\partial \varepsilon_s}
\sigma(\lambda). \label{dos}
\end{equation}

In the absence of magnetic fields, $B\rightarrow\infty$,
the solution of eq.(\ref{sigma}) is expressed as,
\begin{eqnarray}
\sigma(\lambda)=\frac{1}{L}\int^{\infty}_{-\infty}\frac{d\omega}{2\pi}
\frac{e^{u\vert \omega\vert}e^{-i\omega\lambda}}{2\cosh u\omega}
+\int^{\infty}_{-\infty}\frac{d\omega}{2\pi}
\frac{e^{-i\omega\lambda}}{2\cosh u\omega}
\int^{Q}_{-Q}dk\rho(k)e^{i\omega\sin k}. \label{sigds}
\end{eqnarray} 
For $\lambda \gg 1$, the first term of eq.(\ref{sigds}) 
behaves like $\sim 1/\lambda^2$,
while the second term is just the order of $O(e^{-\pi\lambda/u})$.
Thus the main singular contribution comes from the former which is nothing
but the boundary term. 
In a similar manner, from eq.(\ref{enespin}) 
we obtain the asymptotic form of $\varepsilon_s(\lambda)$
for large $\lambda$, i.e. $\varepsilon_s(\lambda)
\sim A e^{-\pi \lambda/u}$, where $A$ is a constant.
Then from eqs.(\ref{dos}) and (\ref{sigds}), we have,
\begin{equation}
\rho_{\rm spin}(\varepsilon)\sim \frac{1}{\varepsilon (\ln\varepsilon)^2},
\end{equation}     
for small $\varepsilon$.
Thus the local density of states also shows the singular divergent
behavior because of the presence of the boundary.
It is noted that this result is similar to 
that of the underscreened multi-channel Kondo effect.\cite{kawa}

The important message of this and the previous subsections is
that in 1D correlated electron systems the localized moment induced 
by a non-magnetic impurity
is not quenched at any temperatures. 
The inclusion of irrelevant interactions such as hopping between
semi-infinite chains does not change the result qualitatively.
It should be stressed that this unquenched local moment is
a particular property of the 1D systems where an impurity divides the system
into two semi-infinite chains.
Such a separation of the system is not possible in higher dimensional
systems.

\subsection{Spin lattice relaxation rate}

Here we calculate the spin lattice relaxation rate $1/T_1$
in the vicinity of a spinless impurity, i.e. a boundary,
using bosonization method and boundary conformal field theory.
The same kinds of the calculations have been done 
for Heisenberg spin chains before.\cite{bru,aff2}
Some parts of the following results are similar to 
those obtained in ref.26 and 27.
However combining them with the results from the Bethe ansatz 
exact solution, we shall see some new aspects. 
In the previous subsections, it was shown that the induced moment 
is not screened completely at any temperatures.
Then one might expect that $1/T_1$ behaves like that of an isolated
spin, $1/T_1\sim T\chi$. However, as will be seen below,
this naive expectation is incorrect.

According to the boundary conformal field theory,
correlation functions for any operators in the vicinity of boundaries
are obtained by the analytic continuation of the antiholomorphic part
to the holomorphic part,
$O(z,\bar{z})=O_L(vt+ix)O_R(vt-ix)\sim O_L(vt+ix)O_L(vt-ix)$.
\cite{boun,boun2}
Following the standard technique, we have the asymptotic
behaviors of the spin-spin correlation function 
in the presence of the boundary,\cite{boun,egg,egg2} 
\begin{eqnarray}
&&\chi(x,y,t)\sim \biggl(\frac{\pi T}{v_s}\biggr)^2
[\sinh\frac{\pi T}{v_s}(x-y-vt)]^{-2} \nonumber \\
&&+e^{2ik_{\rm F}(x-y)}
\prod_{\nu=s,c}
\biggl[\biggl(\frac{\pi T}{v_{\nu}}\biggr)^2
\frac{\sinh\frac{2\pi Tx}{v_{\nu}}\sinh\frac{2\pi Ty}{v_{\nu}}}
{\sinh\frac{\pi T}{v_{\nu}}(x+y+v_{\nu}t)
\sinh\frac{\pi T}{v_{\nu}}(x+y-v_{\nu}t)
\sinh\frac{\pi T}{v_{\nu}}(x-y+v_{\nu}t)
\sinh\frac{\pi T}{v_{\nu}}(x-y-v_{\nu}t)}
\biggr]^{\frac{K_{\nu}}{2}}. \label{spincor}
\end{eqnarray} 
Here $v_s$ and $v_c$ are the velocities of spinon and holon, respectively.
$K_c$ is the Luttinger liquid parameter in the charge sector,
and $1/2 \leq K_c \leq 1$. 
$K_s=1$ because of the $SU(2)$ symmetry of the spin sector.
In the vicinity of the boundary, i.e. $x,y,\vert x-y\vert \ll v_st$,
the staggered part (the second term) of eq.(\ref{spincor})
is less relevant in comparison with the uniform part (the first term).
Thus in contrast to the bulk behavior, 
the antiferromagnetic spin fluctuation is much suppressed,
 and the uniform part gives the dominant contribution to $1/T_1$
near the boundary.
Fourier transforming eq.(\ref{spincor}),\cite{schu,sac} we obtain,
up to logarithmic corrections,
\begin{equation}
\frac{1}{T_1T}=\lim_{\omega\rightarrow 0}\frac{1}{\omega}
\sum_{q,q'}{\rm Im}\chi(q,q',\omega)\sim \frac{C}{v_s^2}+O(T^{2K_c}).
\end{equation}
Here we have assumed that the hyperfine coupling constant 
is independent of $q$, and omitted it.
$C$ is a temperature-independent constant.
In the case that $v_s$ is a constant, the above result is equivalent to
that obtained by Brunel et al. for Luttinger liquids 
with boundaries.\cite{bru} 
The spinon velocity $v_s$ is related to the spin susceptibility
obtained before, $1/v_s=\chi_{\rm bulk}+\chi_{\rm boundary}/L$. 
As claimed in the previous subsections, 
$\chi_{\rm boundary}$ should show enhanced 
local correlations like $\chi_{\rm boundary}\sim 1/T(\ln T)^2$.
Thus near the boundary, 
\begin{equation}
\frac{1}{T_1T} \sim (\chi_{\rm boundary})^2. \label{t1t}
\end{equation}
Surprisingly, this relation is analogous to the Korringa relation.
However it is noted that in contrast to the conventional Korringa law, 
the right-hand side of eq.(\ref{t1t}) shows strong temperature dependence.
As mentioned in the previous subsections, 
the induced local moment is not quenched completely. 
In spite of such an unscreened character of the moment,
{\it a la} Korringa relation holds in the vicinity of a spinless impurity.

\section{A spinless impurity in the 2D and 3D Hubbard models}

In this section, we discuss the local magnetic properties
caused by a spinless impurity in the 2D and 3D Hubbard model 
in the presence of bulk antiferromagnetic fluctuations, i.e.
very close to the half-filling.
The main purpose of this section is to derive the Korringa relation
satisfied at the nearest neighbor of the impurity site, which is
observed in the NMR experiment for cuprates.\cite{nmr4}	
The model hamiltonian is given by,
\begin{equation}
H=\sum_{k\sigma}E_kc^{\dagger}_{\sigma k}c_{\sigma k}
+U\sum_i n_{\uparrow i}n_{\downarrow i}+E_0\sum_{\sigma}n_{\sigma 0}.
\end{equation}
Here $E_k=-2t\sum_{a=1}^{D}\cos k_a$ ($D=2$ or $3$), and 
the last term represents a spinless impurity localized at site $0$.
We restrict our discussion to the case of the square lattice in 2D 
and the cubic lattice in 3D.
Because of the presence of an impurity, correlation functions 
are non-local.  
In the case of $U=0$, the single particle Green's function is given by,
\begin{equation}
G^0_{k k'}(\varepsilon_n)=\frac{\delta_{k k'}}{i\varepsilon_n+\mu-E_k}
+\frac{1}{i\varepsilon_n+\mu-E_k}\cdot
\frac{E_0}{1-E_0\sum_{k''}\frac{1}{i\varepsilon_n+\mu-E_{k''}}}
\cdot\frac{1}{i\varepsilon_n+\mu-E_{k'}}, \label{green}
\end{equation}
where $\mu$ is a chemical potential.
For $U\neq 0$, the single particle Green's function is obtained by
solving the equation,
\begin{equation}
\sum_{k''}[(i\varepsilon_n+\mu-\varepsilon_k)\delta_{k k''}
-\Sigma_{k k''}(\varepsilon_n)-E_0]G_{k''k'}(\varepsilon_n)=\delta_{k k'}.
\end{equation} 
The self-energy $\Sigma_{k k'}(\varepsilon)$ may be obtained
by perturbative calculation in terms of $U$.
However, in the following qualitative argument we do not need 
the explicit expression of $G_{k k'}(\varepsilon_n)$.

Before discussing about magnetic properties, it is useful to 
sketch the spatial dependence of the density of states 
in the vicinity of an impurity.
The density of states at the Fermi level is given by,
\begin{equation}
\rho(x,x')=-(1/\pi)\sum_{k k'}{\rm Im}G_{kk'}^R(0)
e^{ikx}e^{-ik'x'}.
\end{equation}
To simplify the calculation, we consider the strong limit of 
an impurity potential, i.e. $E_0 \gg U, t $.  
The following argument do not change qualitatively
even in the case of a finite $E_0$.
Using eq.(\ref{green}), we obtain the density of states at the Fermi level
for the non-interacting system, $U=0$,
\begin{eqnarray}
\rho_{0}(x,x')=N_{x-x'}(\mu)-\frac{N_{x}(\mu)N_{x'}(\mu)}{N_0(\mu)},
\label{dens}
\end{eqnarray}
\begin{equation}
N_x(\varepsilon)\equiv\sum_k\delta(\varepsilon -E_k)e^{ikx}
=\int^{\infty}_{-\infty}\frac{ds}{2\pi}e^{is\varepsilon}
\prod_{i=1}^{D} J_{n_i}(ts), \label{nx}
\end{equation} 
with $J_n(x)$, the Bessel function, and $x=(n_1,n_2)$ for $D=2$ and
$x=(n_1,n_2,n_3)$ for $D=3$.
Note that if $x$ or $x'$ is the impurity site, the density of states
vanishes, $\rho_{0}(x,0)=\rho_{0}(0,x')=0$.
We can easily show that 
if the electron density is close to the half-filling,
i.e. $\vert\mu\vert/t\ll 1$, then $N_x(\mu)\sim O((\mu/t)^2)$ for
the site $x$ on the sublattice which includes the nearest neighbor
site of the impurity, $x_{\rm n.n.}$ 
(denoted by $A$-sublattice),
and $N_x(\mu)\sim N_0(\mu)$ for the site $x$ on the sublattice
which includes the impurity site ($B$-sublattice).
Thus from eq.(\ref{dens}) we immediately see that the local density of states 
around the impurity site shows strong spatial modulation
similar to the Friedel oscillation.
The period of the oscillation is $\sim 1/k_{\rm F}$ which is close to
the half-filling value in this case.
If $x$ and $x'$ belong to $A$-sublattice, 
the local density of states is nearly equal to that of bulk systems, 
$\rho_0(x,x')\sim 
\rho_0(\vert x\vert\rightarrow \infty, \vert x'\vert\rightarrow\infty)$. 
On the other hand, if $x\neq 0$ and $x'\neq 0$ belong to $B$-sublattice,
$\rho_0(x,x')\sim O((\mu/t)^2)$. 
Since the density of states at the Fermi level is not renormalized by
electron-electron interaction,
this Friedel oscillation occurs for $U\neq 0$.
This observation leads to an important implication for 
local magnetic properties on the nearest neighbor site of the impurity,
$x_{\rm n.n.}$. 
Because of the Friedel oscillation and the bipartite lattice structure,
the local density of states on all the sites surrounding 
the site $x_{\rm n.n.}$ is much suppressed provided that
the impurity potential is sufficiently strong.
Thus the spin on the site $x_{\rm n.n.}$ is less screened
than spins of electrons in the bulk.
As a result, the local spin susceptibility on the site $x_{\rm n.n.}$
is strongly enhanced.

Now we consider the spin lattice relaxation rate $1/T_1$
in the vicinity of a spinless impurity. 
For simplicity, we assume that the hyperfine coupling constant does not
depend on $q$.
We apply the general argument from the Fermi liquid theory
to the case with a single spinless impurity.\cite{koh,shiba}
Then $1/T_1$ at the site $x_i$ is given by, up to constant factors,
\begin{eqnarray}
\frac{1}{T_1T}&=&
\lim_{\omega\rightarrow 0}\sum_{q,q'}\frac{{\rm Im}\chi(q,q',\omega)}{\omega}
e^{iqx_i}e^{-iq'x_i}\nonumber \\
&=&\sum_{q_1,q_2,q_3,q_4,k,k'}{\rm Re}\Lambda(q_1,k+q_2,k)
{\rm Im}G_{k k'}^R(0){\rm Im}G_{k+q_2 k'+q_3}^R(0)
{\rm Re}\Lambda(q_4,k'+q_3,k')e^{iq_1x_i}e^{-iq_4x_i}, \label{t1}
\end{eqnarray}
where $\Lambda(q,k+q',k)$ is a three point vertex function.
The diagrammatic expression of eq.(\ref{t1}) which is the $\omega$-linear
term of ${\rm Im}\chi(q,q',\omega)$ is shown in FIG.1.
The detail derivation of this formula is given in ref.33 and 34. 
In the presence of strong antiferromagnetic fluctuations,
it is plausible to assume that 
$\Lambda(q,k+q',k)$ depends mainly on $q$ and $q'$.
Thus in the following we discard the $k$-dependence of $\Lambda(q,k+q',k)$.
It is useful to rewrite eq.(\ref{t1}) in terms of quantities 
in the coordinate space,
\begin{equation}
\frac{1}{T_1T}=\pi^2 \sum_{s,t}\rho(-x_i-s, -x_i+t)\rho(x_i+s, x_i-t)
\Lambda(x_i,x_i+s)\Lambda(x_i-t,x_i). \label{t12}
\end{equation}
Here,
\begin{equation}
\Lambda(x,x')=\sum_{q,q'}{\rm Re}\Lambda(q,q')e^{iqx}e^{-iq'x'}.
\end{equation}
In the case that the site $x_i$ is far from the impurity,
i.e. $\vert x_i \vert \gg a$, where $a$ is a lattice constant,
$\rho(x_i+s,x_i-t)\rightarrow \rho(s+t)$, 
$\Lambda(x_i,x_i+s)\rightarrow \Lambda(-s)$, and thus 
the above expression is reduced to the usual formula of
$1/T_1$ in bulk systems.
It is also noted that if $x_i=0$, $1/T_1T$ vanishes, since
$\Lambda(0,s)$ includes $G(0,x_j)$ which vanishes as mentioned above.
Here we are concerned with the case that 
the site $x_i$ is the nearest neighbor of the impurity site, 
$x_i=x_{\rm n.n.}$.

To proceed further, 
we use a phenomenological expression for $\Lambda(q,q')$. 
We assume that the three point vertex function $\Lambda(q,q')$ 
consists of the part which is strongly enhanced by 
antiferromagnetic spin fluctuation and the local part
which depends on $q$ and $q'$ weakly,
\begin{equation}
\Lambda(q,q')\sim \Lambda_{\rm AF}(q)\delta_{q,q'}+\Lambda_{\rm loc}(q,q').
\end{equation}
The antiferromanetically correlated part $\Lambda_{\rm AF}(q)$ has 
a strong peak at $q=Q$, the staggered vector, and then approximated as,
${\rm Re}\Lambda_{\rm AF}(q)\sim {\rm Re}\chi(q\sim Q)
=\chi(Q)/(1+(\xi_{\rm AF}(q-Q))^2)$.
Here we used the phenomenological expression for $\chi(q\sim Q)$.\cite{moriya}
As mentioned above,
$\Lambda_{\rm loc}(q,q')$ is enhanced by local magnetic correlations at
the site $x_{\rm n.n.}$, i.e. 
$\Lambda_{\rm loc}(q,q')\sim \Lambda_{\rm loc}
e^{-iqx_{\rm n.n.}}e^{iq'x_{\rm n.n.}}$
Then, eq.(\ref{t1}) is rewritten into,
\begin{eqnarray}
\biggl(\frac{1}{T_1T}\biggr)_{\rm n.n.}&\sim &
\biggl[\frac{\chi(Q)}{(\xi_{\rm AF})^m}\biggr]^2
\sum_{k,k'}{\rm Im}G_{k,k'}^R(0){\rm Im}G_{k+Q,k'+Q}^R(0) \nonumber \\
&&+[{\rm Re}\Lambda_{\rm loc}]^2
\sum_{k,k',q_2,q_3}{\rm Im}G_{k,k',q_2,q_3}^R(0){\rm Im}G_{k+q_2,k'+q_3}^R(0).
\label{t13}
\end{eqnarray}
where $m=2$ for 2D systems and $m=3$ for 3D systems.
Since $\chi(Q)\sim (\xi_{\rm AF})^2$,\cite{moriya}
the first term of eq.(\ref{t13}), which is 
the antiferromagnetically correlated part, is much suppressed 
compared to the second term, i.e. the local correlation part.
Thus we obtain,
\begin{equation}
\biggl(\frac{1}{T_1T}\biggr)_{\rm n.n.}
\sim ({\rm Re}\Lambda_{\rm loc})^2. \label{t14} 
\end{equation}
Here we neglect all factors which are not enhanced by electron correlation. 
On the other hand, the local spin susceptibility at $x_{\rm n.n.}$
is approximately given by, $\chi_{\rm loc}\sim {\rm Re}\Lambda_{\rm loc}$.
Thus eq.(\ref{t14}) establishes the Korringa relation satisfied
at the nearest-neighbor site of the impurity. 
As mentioned before, this relation is actually 
observed in NMR experiments.\cite{nmr4}

\begin{figure}
\centerline{\epsfxsize=8cm \epsfbox{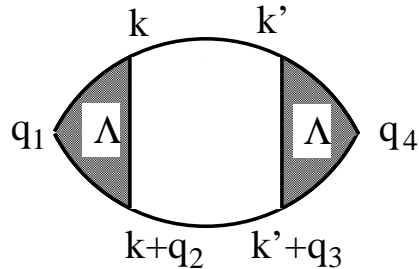}}
\caption{diagram of $\omega$-linear term of ${\rm Im}\chi(q_1,q_4,\omega)$.
The shaded part is the three point vertex.}
\end{figure}

\section{Summary and Discussion}

We have discussed some magnetic properties analogous to the Kondo effect
induced by a spinless impurity
in strongly correlated electrons systems.
In the 1D system, we have shown that the spin susceptibility 
and the local density of states near the impurity indicate 
divergent behaviors implying the presence of an unquenched local moment
at any temperatures.
We have also obtained the Korringa-like relation between 
the spin lattice relaxation rate and the local spin susceptibility,
$1/T_1T\sim(\chi_{\rm boundary})^2$.
In the 2D and 3D systems, the antiferromagnetically correlated
Fermi liquid theory has been applied.
It has been shown that magnetic properties in the vicinity of 
a spinless impurity are dominated by the induced moment rather than
the antiferromagnetic spin fluctuation developed
in the bulk, and that the Korringa relation holds at the 
near neighbor site of the impurity.

The results obtained for the 1D system have an interesting
implication to higher dimensional systems.
Suppose a semi-infinite 2D Hubbard model with a boundary line,
which is regarded as the coupled semi-infinite Hubbard chains.
According to the results obtained in Sec. II, it is expected that
at some finite temperatures the 1D-like strong spin correlations occurs
in the vicinity of the boundary line leading to strongly enhanced 
density of states near the boundary.
Such an enhanced electron correlation and one dimensionality 
of the boundary line may give rise strong fluctuations toward
some surface phase transition.
For instance, if there exists a pairing interaction in the bulk system,
the paring correlation may be enhanced near the boundary leading to
higher transition temperature than the bulk superconductivity.
Actually, it is reported that ${\rm Sr_2RuO_4}$ with lamellar
microdomains of Ru metal shows the superconducting
transition at the temperature higher than $T_c$ of the pure system, 
and that the superconductivity with higher-$T_c$ occurs in the vicinity
of the boundary between ${\rm Sr_2RuO_4}$ and Ru-metal.\cite{maeno}
We would like to pursue this possible mechanism of the enhanced transition
temperature in the near future. 

\acknowledgments{}
The author thanks N. Kawakami, H. Frahm, and K. Yamada 
for invaluable discussions.
This work was partly supported by a Grant-in-Aid from the Ministry
of Education, Science, Sports and Culture, Japan.


                                                                    
\end{document}